\documentclass[american, 10pt]{article}

\usepackage{amsfonts}
\usepackage{amsmath}
\usepackage{amssymb}
\usepackage{babel}
\usepackage[a4paper, top=1.2cm, bottom=1.2cm, left=2.7cm, right=2.7cm,
            includefoot, includehead]{geometry}
\usepackage{graphicx}
\usepackage{flafter}
\usepackage[figurename=Fig., labelsep=period, labelfont=bf,
            format=hang, justification=raggedright]{caption}
\usepackage[bookmarks=true, bookmarksnumbered=true,
            bookmarksopen=false, bookmarksopenlevel=2,
            pdfborder={0 0 1}, breaklinks=false,
            backref=false, colorlinks=false]{hyperref}
\hypersetup{pdftitle = {Emergence from Symmetry: A New Type of Cellular Automata},
            pdfauthor = {Zan Pan},
            pdfsubject = {Cellular Automata},
            pdfkeywords = {cellular automaton, symmetry, emergence, Game of Life}}
\graphicspath{{figures/}}

\begin{document}
\title{\textbf{Emergence from Symmetry:\\A New Type of Cellular Automata}}
\author{Zan Pan
 \footnote{E-mail address:
           \href{mailto:panzan@mail.ustc.edu.cn}{\tt{panzan@mail.ustc.edu.cn}}}\\
 \textsc{\small Department of Modern Physics,}\\
 \textsc{\small University of Science and Technology of China, Hefei, 230026, P. R. China}}
\date{}
\maketitle

\begin{abstract}
 In the realm of cellular automata (CA), Conway's Game of Life (Life) has gained the most
 fondness, due to its striking simplicity of rules but an impressive diversity of behavior.
 Based on it, a large family of models were investigated, e.g. Seeds, Replicator, Larger
 than Life, etc. They all inherit key ideas from Life, determining a cell's state in the
 next generation by counting the number of its current living neighbors. In this paper,
 a different perspective of constructing the CA models is proposed. Its kernel, the Local
 Symmetric Distribution Principle, relates to some fundamental concepts in physics,
 which maybe raise a wide interest. With a rich palette of configurations, this model
 also hints its capability of universal computation. Moreover, it possesses a general
 tendency to evolve towards certain symmetrical directions. Some illustrative examples
 are given and physical interpretations are discussed in depth. To measure the evolution's
 fluctuating between order and chaos, three parameters are introduced, i.e. order parameter,
 complexity index and entropy. In addition, we focus on some particular simulations and
 giving a brief list of open problems as well.
\end{abstract}

\section{Introduction}
 The conception of cellular automaton (CA) stems from John~von~Neumann's research on
 self-replicating artificial systems with computational universality~\cite{NB}. However,
 it was John~Horton~Conway's fascinating Game of Life (Life)~\cite{Gardner, BCG} which
 simplified von~Neumann's ideas that greatly enlarged its influence among scientists.
 It vividly illustrates the emergence of complex behavior from simple rules and thus
 possesses profound philosophical connotations. Albeit useless for computation in
 practice, Life has been proven theoretically as powerful as a universal Turing
 machine~\cite{BCG}. Since then, CA has widely spread its voice outside computer
 laboratories. A plethora of models are developed to simulate physical or social
 systems, such as lattice gas automata in molecular dynamics~\cite{HPP, FHP, Meyer},
 the Asymmetric Simple Exclusion Process on traffic problems~\cite{Schadschneider} and
 Langton's ant--- a particular species of artificial life~\cite{Langton}. For those
 who are interested, Ref.~\cite{CD} provides a detailed discussion of various CA models.
 Besides, comments on CA together with other extremely simplified models in physics can
 be found in~\cite{Bagnoli}. A thorough survey on the aspects concerning mathematical
 physics refers to~\cite{Smith}.

 To trace the development of physical thoughts in this field, we would like to
 mention two books. In the late 1970s, Konrad~Zuse conceived an essay entitled
 \textsl{Calculating Space}~\cite{Zuse}, in which he advocated that physical laws
 are discrete by nature and that the entire history of our universe is just the
 output of a giant deterministic CA. Although crazy-sounding, it indeed suggested
 a possible replacement of traditional depicts in our textbooks, and signified the
 outset of \emph{digital physics}. Thirty-three years later, Stephen~Wolfram, a pioneer
 of researches on the complexity of CA~\cite{Wolfram83, Wolfram84}, published a notable
 book named \textsl{A New Kind of Science}~\cite{Wolfram02}, in which he extensively
 and systematically argued the discoveries on CA are not isolated facts but closely
 interrelated with all kinds of disciplines.

 No doubt that the invention of Life marked a watershed in the history of CA. It has
 developed a lasting cult and has many variations (see~\cite{Eppstein} and references
 therein). In the compact notation used by \textsf{Golly}\footnote{\label{Golly}
 \textsf{Golly} is a cross-platform open-source simulation system for Life and some
 other interesting CA models. It is available at \href{http://golly.sourceforge.net/}{\texttt{http://golly.sourceforge.net/}}.},
 Life is denoted as B3/S23, where ``B'' stands for ``birth'' and ``S'' for ``survival''.
 In a nutshell, Life is mainly generalized through three ways. (i)\;Simply change the
 number of living neighbors that determines dead cells' birth and living cells' survival.
 Some examples include Replicator (B1357/S1357), Seeds (B2/S) and Plow World (B378/S012345678).
 (ii)\;Assign more states to a cell or modify the geometry of the universe. One of the
 most famous is a three-state (live, ghost, death) automaton called Brian's Brains.
 (iii)\;Extend the distance of Moore neighborhood beyond one, such as Larger than
 Life~\cite{Evans} and Kaleidoscope of Life~\cite{ALPU}. In the latter, a cell's state
 is determined by whether there are exactly 4 living cells at Moore-distance 1 or 2.
 Since Life is at the root of these models, they all share certain sameness inevitably.
 To explore something utterly different from Life, novel methods to construct CA models
 are needed.

 We will begin in Sec.~\ref{symmetry} with introducing the transition rules, giving some
 examples and outlining its features. Then, physical interpretations of the automaton are
 viewed from different perspectives (Sec.~\ref{interpretation}). It should be noted that
 those discussions can not be limited to this model only. In Sec.~\ref{parameters}, three
 evolution parameters are defined and discussed. Despite that insufficient efforts are
 spent on these parameters (no specific assertions are made), they may exemplify a heuristic
 approach to mathematical investigations. After taking an especial look at several
 independent topics in Sec.~\ref{exploration}, the paper finishes with a brief summary
 and a reemphasis on Zuse's paradigm-shifting ideas (Sec.~\ref{conclusion}).

\section{Game of Symmetry\label{symmetry}}
\subsection{Rules}
\begin{figure}[h]
 \centering
 \includegraphics[width=420 pt]{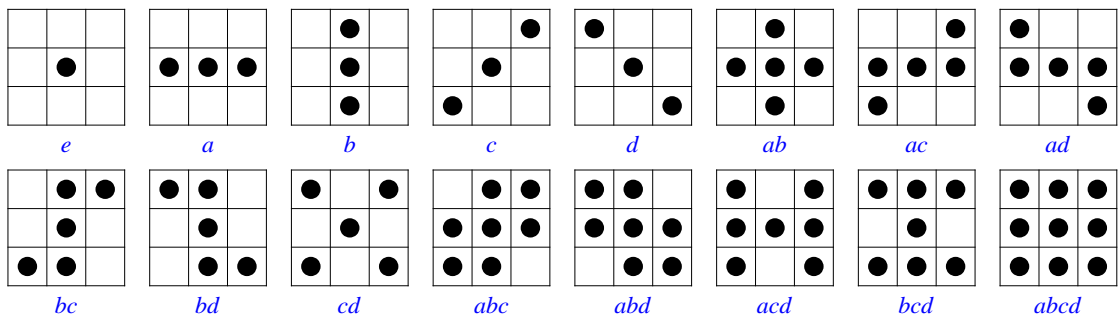}
 \caption{\label{meta}Sixteen meta-configurations}
\end{figure}

 As the same to Life, we adopt Moore neighbors to construct this CA model; that is,
 a neighborhood consists of nine cells. Each cell has two possible states: dead and
 alive. For the ease of formulation, sometimes we may use empty and occupied, or 0
 and 1 instead if convenient. For intuition, we also use particles to indicate living
 cells, generating and annihilating to describe the evolution process of birth and
 death, and so forth. The rules we will set down are called \emph{the Local Symmetric
 Distribution Principle} (LSDP). Among 512 possible patterns for a neighborhood, sixteen
 are considered as \emph{meta-configurations} (Fig.~\ref{meta}), the distribution of
 whose Moore neighbors is symmetrical with respect to the center particle. Actually,
 the first five comprise all the rest via a special superposition. More precisely, the
 set of those configurations forms a bounded semilattice $\mathcal{S}$. In other words,
 it is a commutative monoid with the presentation
\begin{equation}
 \mathcal{S}=\langle a, b, c, d\,|\,a^2=a,\,b^2=b,\,c^2=c,\,d^2=d\,\rangle.
\end{equation}
 It is assumed that each neighborhood spontaneously tries to reach one of the
 meta-configurations, i.e. all cells are updated simultaneously, conforming to
 the following transitions:
\begin{enumerate}
 \item A dead cell comes back to life when its neighbors are symmetrically distributed.
 \item A living cell with symmetrically distributed neighbors survives to the next
       generation.
 \item A living cell with asymmetrically distributed neighbors transitions distinctively
 in two cases: if activating one dead neighbor would get to a meta-configuration, it just
 introduces such a desired change; otherwise, it becomes a dead cell.
\end{enumerate}

 This model is also a zero-player game. The evolution is determined by applying the above
 rules repeatedly to each cell in the former generation. Now a cell's state depends on the
 distribution of its neighbors rather than the sum of their states (express the rules in Life
 using $0/1$ notation), therefore it is no longer a totalistic automaton. As a consequence,
 different landscapes in this ``universe'' can be expected. A further understanding of LSDP
 will be discussed in Sec.~\ref{interpretation}, and now we shall see some useful configurations
 to get familiar with the rules.

\subsection{Examples of Patterns}
\begin{figure}[htbp]
 \centering
 \includegraphics[height=90 pt]{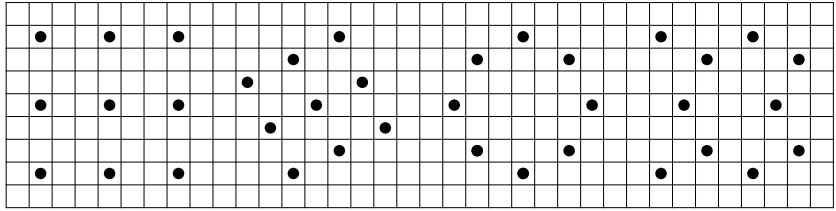}
 \caption{\label{still}Some patterns of still lives in $7\times7$ grid}
\end{figure}

\begin{figure}[htbp]
 \centering
 \includegraphics[width=120 pt, height=120 pt]{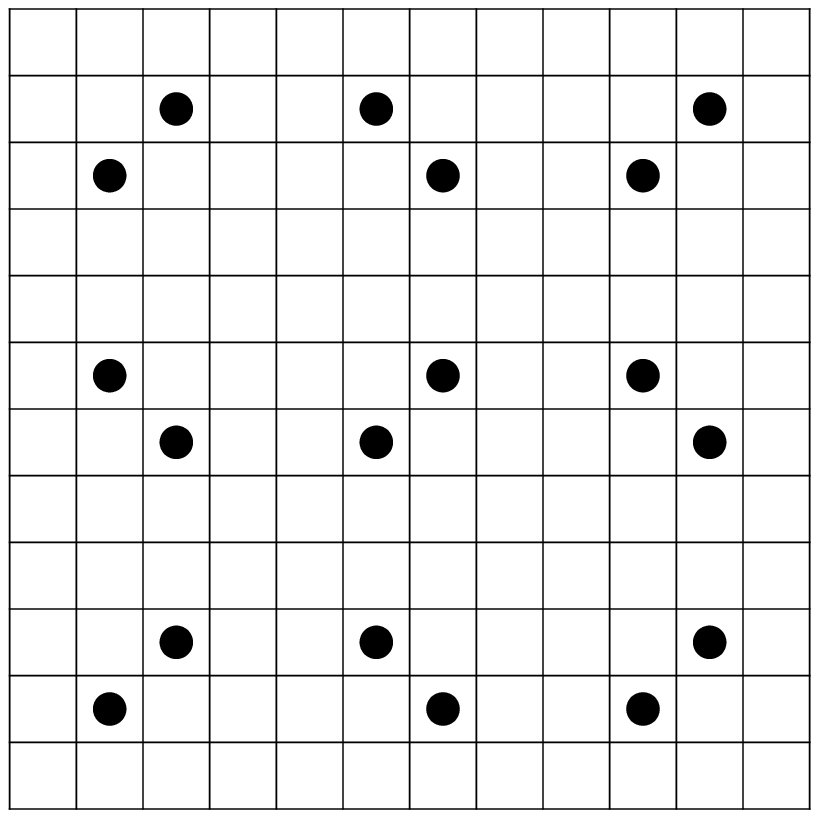}
 \hspace{10 pt}
 \includegraphics[width=160 pt, height=120 pt]{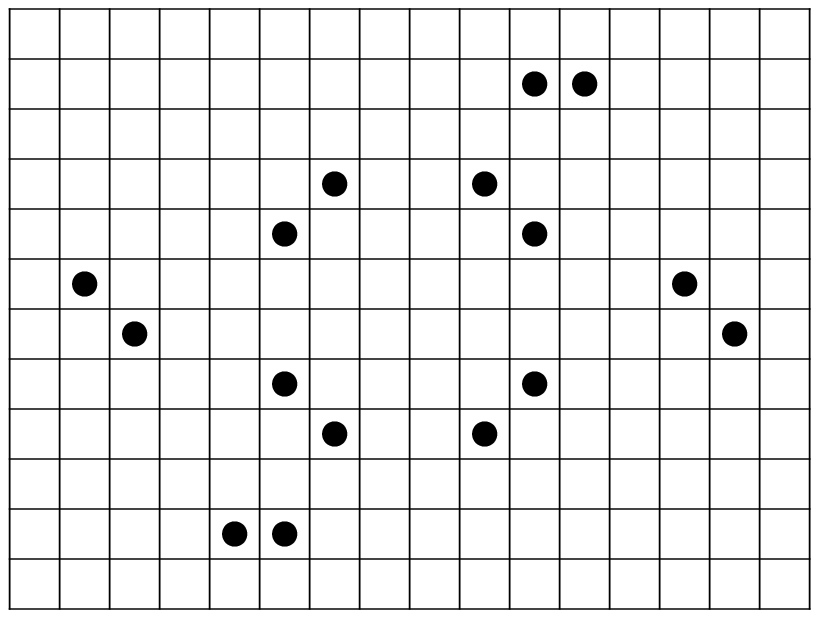}
 \caption{\label{oscillator}Two constructions of particle emitters}
\end{figure}

 The simplest static patterns are \emph{still lives}, including an isolated particle,
 i.e. the first configuration in Fig.~\ref{meta}. It seems that all still lives are
 collections of isolated particles (Fig.~\ref{still}). Experimentations show that the
 largest number of noninteracting particles that can be placed in $7\times7$ grid is
 10. Remarkably, the first two patterns in Fig.~\ref{still} can be used as different
 lattices in the simulations of physical systems. \emph{Oscillators}, the majority of
 which are period 2, can be assembled easily from several basic modules that will be
 classified later. It is worth special notice that the octagonal configuration in
 Fig.~\ref{oscillator} serves as a base of various particle emitters. Of course, there
 exist many other configurations that emit particle beams (particles propagating along
 a line). For instance, the toad configuration\footnote{See
    \href{http://en.wikipedia.org/wiki/Conway's_Game_of_Life}
       {\tt http://en.wikipedia.org/wiki/Conway's\_Game\_of\_Life}.}---an oscillator in
 Life---is also an oscillator in our model. Simply adding one particle in the appropriate
 cells will just obtain an emitter.

\begin{figure}[htbp]
 \centering
 \includegraphics[height=90 pt]{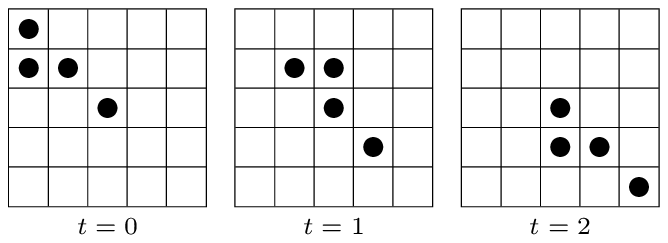}
 \hspace{10 pt}
 \includegraphics[height=90 pt]{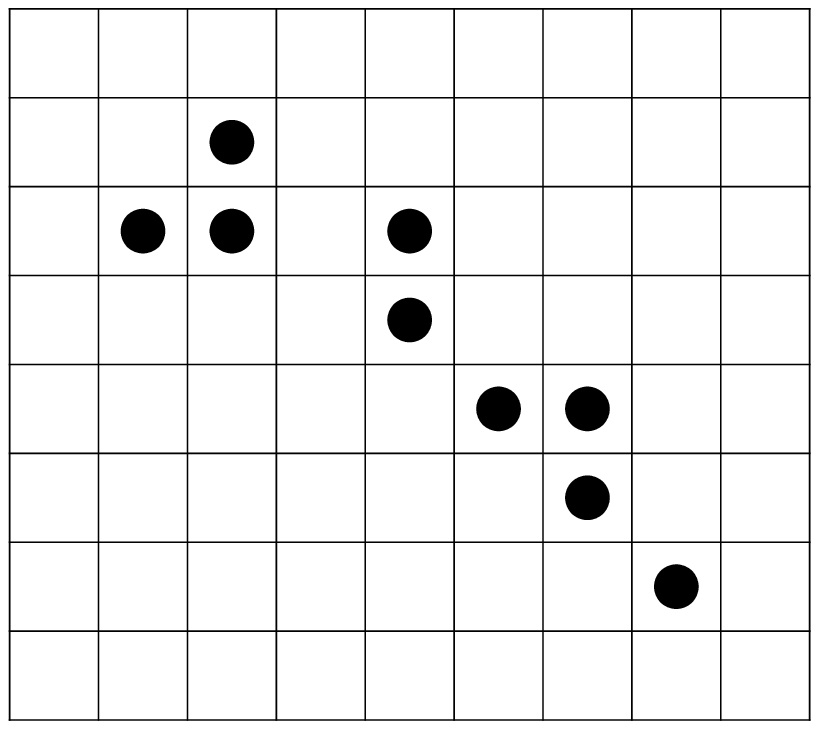}
 \caption{\label{glider}The simplest glider's evolution (left) and a lightweight 
  spaceship (right)}
\end{figure}

 Other funny patterns include \emph{gliders} or \emph{spaceships}, a frequent occurrence
 of which in Life makes it Turing complete. This kind of patterns is the most valuable
 and usually at the heart of a universal constructor. Collision-based computation (see
 Ref.~\cite{Adamatzky}) is so widely studied that perhaps it has set apart a pure research
 field called \emph{glider dynamics}. Two examples are shown in Fig.~\ref{glider}. They
 both have a period of 2 and travel across the grid diagonally at the same velocity.

\begin{figure}[htbp]
 \centering
 \includegraphics[height=75 pt]{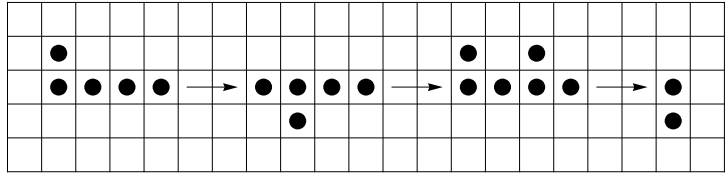}
 \caption{\label{diverter}Evolution of a simple diverter ($n=4$)}
\end{figure}

 Patterns called \emph{diverters} refer to those configurations that can change their
 evolution direction in finite time. An initial state like that in Fig.~\ref{diverter},
 only differing in the number of horizontally distributed particles (denoted by $n$,
 $n\geqslant3$), will eventually extend itself in vertical direction after $n-1$ generations.
 In sharp contrast, this series of configurations in Life nearly have nothing in common
 during evolution, reflecting a highly dependence on the absolute number of neighbors'
 population rather than global distribution. Therefore, this model may not be so
 sensitive to a external disturbance. It is meaningful to reveal how global effects
 emerge from local rules.

\subsection{Features}
 The most striking feature of this automaton is the much easier implementation of infinite
 growth (usually at a slow rate). It needs only two particles, directly adjacent to each
 other horizontally, vertically or diagonally, to form such a pattern. For this advantage,
 more convenience is provided to simulate interactions between relatively independent
 systems and thence more powerful calculating machines could be expected. But on the other
 hand, it becomes more impossible to predict the long-term output of the evolution, not
 merely because of chaotic factors but also the high chances that two ostensibly
 disconnected systems can amalgamate with each other after an extremely long time.

 The key to Life's fascination lies in its achieving an astonishing diversity of behavior,
 fluctuating between randomness and order. However, most interesting patterns in the Life
 lexicon, e.g. \emph{Gosper's glider gun}, \emph{puffer train}, and \emph{Methuselahs} are
 all obtained through brute-force computer experimentations. Game of Symmetry can add some
 design into this method to construct more complicated objects by virtue of its numerous
 and flexible subassemblies. Some have been alluded before and we shall see more in
 Sec.~\ref{collision}.

\begin{figure}[htbp]
 \centering
 \includegraphics[width=220 pt]{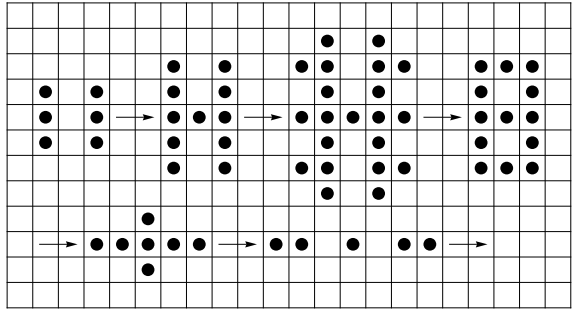}
 \caption{\label{beams}Interactions between two parallel particle beams that illustrate
  how the system's evolution direction changes in order to reach a symmetrical state}
\end{figure}

 Another feature is the general evolution tendency to become symmetrical. Once it happens,
 the symmetry may increase in richness but cannot be lost unless a nearby subpattern comes
 close enough to bring about an unrecoverable interference. This characteristics is also
 possessed by Life and some other models. However, our automaton seems to have a specific
 ability to recognize how the configuration lacks symmetry and automatically evolves
 towards the direction in which symmetry is broken most.  There is no difficulty in
 understanding that such a property have already been embodied in the transition rules.
 This point can be demonstrated by diverters and interactions between parallel particle
 beams (Fig.~\ref{beams}). Although the universe has exhibited a favor of uniformity,
 sometimes it can become totally disordered when such a persistence is impossible. As
 the famous warning puts, \emph{nature does not always share our tastes about a beautiful
 theory}~\cite{Rovelli}, we can not expect the universe always operates according to our
 preferences.

 \subsection{About SuperSymmetry}
 All configurations in this paper have been tested on the program \textsf{SuperSymmetry}
 (it has no much relation to the supersymmetry in string theory). To realize some
 similar features to \textsf{Golly}, such as an infinite universe, extremely fast
 generation and unbounded zooming out for astronomical patterns, it is still under
 development.

\section{Physical Interpretation\label{interpretation}}

 Symmetry, conveying much more meanings than geometrical aesthetics, involves many
 fundamental concepts of modern science. In particle physics, symmetries of the laws
 almost determine the properties of particles found in nature. Noether's renowned
 theorem states that, roughly speaking, for every symmetrical transformation, there
 is a corresponding conserved quantity. Philip Warren Anderson had made a succinct
 comment in his thought-provoking article: \emph{it is only slightly overstating the
 case to say that physics is the study of symmetry}~\cite{Anderson}.

 As already noted, this automaton has an intrinsic property of conserving symmetry
 in the initial configuration, which results from transition rules. Nevertheless,
 the universe still lacks spatial isotropy, although the meta-configurations are
 selected regardless of such a difference. Even from the mathematical angle, LSDP
 is somewhat special. It should be regarded as another type of symmetry that cannot
 be classified into any existing categories. Recall one of the principles in quantum
 mechanics, which posits that  particles of the same kind are utterly indistinguishable.
 To a degree, LSDP is close to Bose-Einstein statistics, allowing two or more identical
 particles to occupy the same state. However, such an analogy is superficial; perhaps
 it associates with a hidden facet of the universe.

 Moreover, systems governed by LSDP tend to evolve towards a pattern in which all
 particles resemble each other from a local perspective, as if a force is exerted.
 This maybe result in the order creation form chaos. According to the contentions
 in Refs.~\cite{PS, Lauglin}, self-organization in a dynamical system is characterized
 by a final state which is more ordered than the initial; and the notion of emergence
 means that the whole is greater than the sum of the parts. Referring to the examples
 and features in Sec.~\ref{symmetry}, we can clearly see that our model is just an
 excellent conceptual laboratory for illustrating these abstract notions (or many
 others).

 Normally, CA models use local rules to determine how a cell transitions, i.e. it only
 considers the nearby neighbors. When designing an algorithm, we never need to make
 calculations over all cells. Whereas, in the real-world, especially at the microscopic
 level, strong-coupling systems are common, such as quantum entanglement and
 superconductivity. Therefore, it appears that our model fails to simulate the whole
 universe. However, if we suppose the automaton evolves extremely fast, e.g. $10^{-43}\;
 \mathrm{s}$ (Planck time) per generation, the local rules will certainly produce the
 same outcomes as nonlocal ones. In this sense, Zuse's suggestion can be quite reasonable,
 provided that the cell is also at the Planck scale ($10^{-35}\;\mathrm{m}$). Remarkably,
 Gerardus~'t~Hooft had expressed similar ideas in~\cite{Plus}. Based on a cogent
 argument, 't~Hooft said: \emph{I think Conway's game of life is the perfect example
 of a toy universe. I like to think that the universe we are in is something like this}.

 Nowadays, most physicists would agree on the statement that \emph{information is
 physical} (Rolf~Landauer's aphorism). Nevertheless, when it comes to the idea that
 \emph{everything is information}\footnote{John Wheeler also made another pithy
 summary of his insights: \emph{It from bit}~\cite{MTZ}. Building upon the researches
 into black hole thermodynamics and the holographic principle (see~\cite{Bousso} for
 reference), Jacob Bekenstein extended Wheeler's original ideas in~\cite{Bekenstein}.},
 controversies are ubiquitous. Notwithstanding that there is not yet a theory with
 information as its core having reached the status of physical law, it is much better
 to leave adequate room for such a belief. We hope that it can blossom in years to
 teach us how to think computationally about nature and bring us fruitful insights on
 physics. Just as the research on automata has implied, the discretization of both space
 and time may be the best clue guiding us to the theory of quantum gravity\footnote{In loop
 quantum gravity (LQG), space-time has a granular structure and each node in a spin
 network determines an elementary grain of space~\cite{Smolin, Ashtekar}. A particular
 perspective to combine digital physics with LQG is discussed in~\cite{Zizzi}.}.

\section{Evolution Parameters\label{parameters}}
\subsection{Symmetry-evolution Diagram}
 For the sake of simplicity, we shall discuss some preliminary terms first. For a
 living cell, its eight neighbors forms \emph{a local environment} of the center
 particle, thus a separate system at most has 256 different local environments.
 We say a particle symmetrical if its environment belongs to one of the sixteen
 meta-configurations; otherwise, it is considered to be asymmetrical. To depict
 the process of order-disorder transitions, it is convenient to use $n_1$, $n_2$ to
 denote the number of symmetrical and asymmetrical particles respectively and then
 define the \emph{order parameter} by the expression
 \begin{equation}
   \xi=\frac{n_1-n_2}{n_1+n_2}.
 \end{equation}
 Order parameter is a term from the theory of phase transition and critical phenomena.
 However, it has completely different meanings here. Obviously, $\xi=1$ relates to a
 still-live pattern; and a dynamically evolving configuration may get closer and closer
 to 1 but can never reach. It is worthy of attention that $\xi=-1$ doesn't mean the
 corresponding configuration will disappear or break down soon. On the contrary, such
 a pattern as a whole can be quite stable (see Fig.~\ref{glider}).

\begin{figure}[htbp]
  \centering
  \includegraphics[height=170 pt]{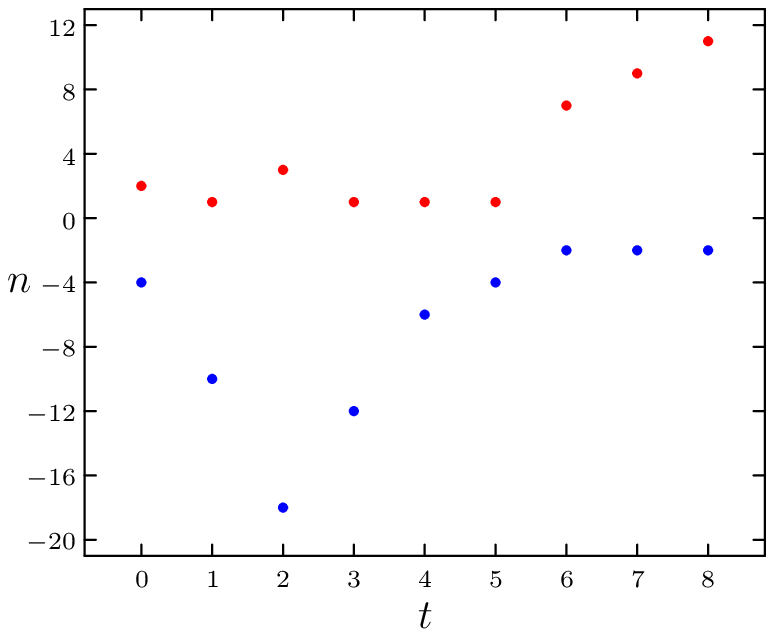}
  \hspace{4pt}
  \includegraphics[height=170 pt]{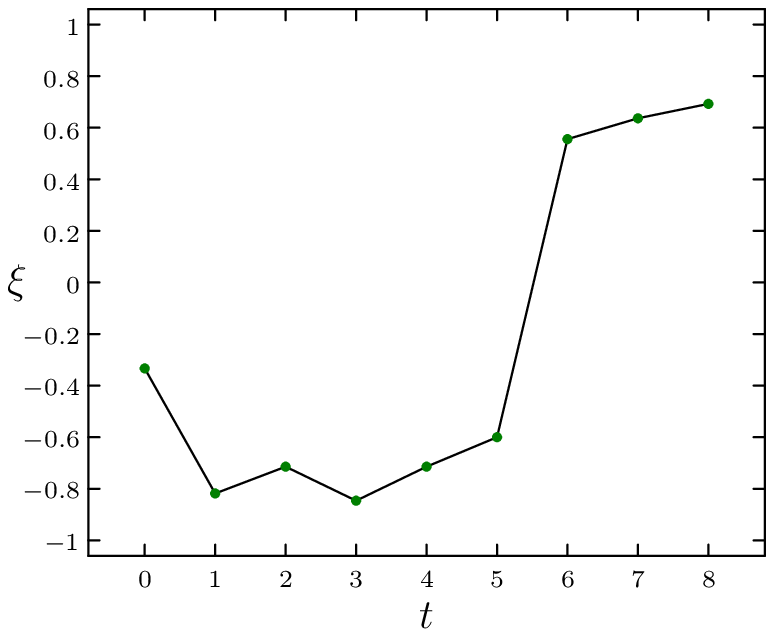}
  \caption{\label{parameter}Particle number of the beam configuration in Fig.~\ref{beams}
   fluctuates irregularly before it reaches convergence to a quite stable state (left).
   The corresponding order parameter curve is also given (right).}
 \end{figure}

 In addition, \emph{symmetry-evolution diagrams} are used for graphical representations.
 In Fig.~\ref{parameter}, we plot the particle number $n$ against evolution time $t$.
 If positive, $n$ represents symmetrical particles; otherwise, it indicates asymmetrical
 ones. This convention does not hold in Table~\ref{data}, where $n=n_1+n_2$. In some
 cases, this kind of diagrams can be extremely irregular, which may provides a possible
 usage of the particle numbers as pseudo-random numbers.

\subsection{Complexity Index}
 Since a pattern can easily grow to infinity, the comparison of two configurations'
 complexity cannot merely rely on their absolute population number. \emph{Complexity
 index} $\phi (t)$ is defined as a function referring to the total number of different
 local environments in a generation. For example, the complexity index of the simplest
 glider (Fig.~\ref{glider}) is invariant during its evolution, i.e. $\phi(t)=4$, $\forall
 \,t>0$. It is evident that all still-live patterns have the same complexity index of 1,
 thus they belong to the same \emph{complexity class}. A larger index usually corresponds
 to more complex behavior. However, this term invites the problem of how to deal with
 separated parts that coexist but have no impacts on each other. Different means to do
 this will yield different values. In principle, it depends on whether you would like
 to regard those parts as mutually independent closed systems or just subsystems of a
 larger one.

\subsection{Entropy}
 We introduce the concept of \emph{entropy} for each generation by adopting Shannon's
 definition of classical information entropy:
 \begin{equation}
   \mathcal{H}=-\sum_{i}p_i\log_{2}{p_i},
 \end{equation}
 where $p_i, i=1,2,\dots,$ are proportional distributions of particles' local environments.
 Meanwhile, the symbol $\mathcal{P}$ is used to denote the \emph{maximum probability} in this
 distribution. We still take the pattern in Fig.~\ref{beams} as an example to illustrate
 the calculations. From Table~\ref{data}, we can conclude that entropy decreases as the
 configuration goes to a more ordered specification in the next generation. By contrast,
 the thermal entropy of any isolated systems must increase, or remains constant at best.
 However, this remark does not connote that our model forbids any continuous increment
 of entropy. In a subtle way, entropy relates to both particle numbers and the complexity
 index, but it is much more conducive to make global evaluations. Due to its far-reaching
 implications, discussions about the concept of entropy can also be found in many surveys
 on CA. Ref.~\cite{BR} reveals a simple linear relation between entropy and the largest
 Lyapunov exponent in the lattice gas model, which may be enlightening.
 \begin{table}[htbp]
 \centering
 \caption{\label{data}Evolution parameters of the configuration in Fig.~\ref{beams}}
 \setlength{\tabcolsep}{4mm}
 \setlength{\arrayrulewidth}{0.05mm}
 \setlength{\doublerulesep}{1mm}
 \renewcommand{\arraystretch}{1.3}
 \begin{tabular}{|c|c|c|c|c||c|c|c|c|c|}
  \hline
  $t$ & $n$ & $\phi$ & $\mathcal{H}$ & $\mathcal{P}$ &
  $t$ & $n$ & $\phi$ & $\mathcal{H}$ & $\mathcal{P}$ \\
  \hline \hline
  0 &  6 &  3 & 1.5850 & 0.3333 & 4 &  7 & 7 & 2.8074 & 0.1429 \\
  1 & 11 &  9 & 3.0958 & 0.1818 & 5 &  5 & 5 & 2.3219 & 0.2000 \\
  2 & 21 & 21 & 4.3923 & 0.0476 & 6 &  9 & 3 & 0.9864 & 0.7777 \\
  3 & 13 & 11 & 3.3927 & 0.1538 & 7 & 11 & 3 & 0.8659 & 0.8182 \\
  \hline
 \end{tabular}
 \end{table}

\section{Further Exploration\label{exploration}}
\subsection{Particle Collision\label{collision}}
\begin{figure}[htbp]
 \centering
 \includegraphics[width=420 pt]{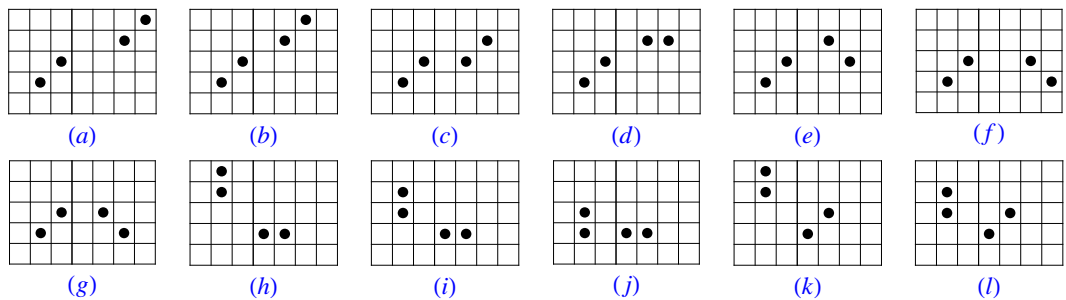}
 \caption{\label{collisions}Collisions between two simplified beams}
\end{figure}

 Particle-like structures in CA do not gain sufficient attention among
 researchers\footnote{We refer to~\cite{BNR} for a discussion of such structures
 in one-dimensional automaton.}. Perhaps this is due to the lack of an ideal model.
 Fortunately, our automaton offers a dozen of patterns facilitating such an investigation.
 Here we only consider the interactions between two particle beams that travel towards
 each other. To simplify our description, the number of particles in each beam is restricted
 to 2 and configurations are set so that they will immediately start interactions in the
 next generation. Fig.~\ref{collisions} shows a few examples. Elaborate arrangements of
 $(a)$, $(c)$, $(d)$, $(f)$ and $(l)$ can constitute many beautiful oscillators. The
 configuration in $(g)$ provides us a neat example to illustrate the conservation of
 momentum and the generation of two gliders. In contrast, $(h)$ acts quite strangely: it
 creates isolate particles along a diagonal, i.e. the symmetry axis. Besides, $(b)$, $(e)$
 and $(k)$ also display gliders' creation from collisions between particle beams, in which
 $(b)$ only generates the simplest gliders while $(k)$ can also produce the lightweight
 spaceship in Fig.~\ref{glider}. All these are resources of great utility, which can be
 used as basic modules when designing complicated projects on this automaton. Some of
 them may play crucial roles in universal computation, such as transporting data,
 self-replicating and generating signal sequences. However, it should be pointed out
 that the evolution may be slightly different in certain patterns when beams consist
 of more particles than two.

\subsection{Evolution from Randomness}
 Quantum field theory tells us that the vacuum is not empty and that all fields undergo
 quantum fluctuations, leading to particle pair production and annihilation all the time.
 Albeit counter-intuitive, this kind of picture has profoundly enhanced our understanding
 of nature. Thus, the evolution results from random initial configurations also deserve a
 big effort. They are essential to evaluate the thesis whether physical laws can emerge
 from random processes, just as John Wheeler puts it: \emph{Law without Law}~\cite{MTZ}.
 Besides, they are helpful to give a comprehensive impression of this model's hallmark.

 If we only consider a cell's nearest neighbors, taking no account of those at Moore
 distance 2 (neighbors of $d \geqslant 3$ doesn't contribute to its fate in the next
 generation), the probability of a cell's being alive will be $31/512\approx 0.06$.
 It is a small value, but the strong interactions between adjacent particles
 may endow them a much longer life, as manifested in the glider's evolution. Therefore,
 we should feel confident of the richness emerging from random specifications.

\subsection{Open Problems}
 It is often hard to cover many aspects in a single manuscript. The following problems
 are selected with the criteria of its significance and the fascination to pique
 widespread interest:
\begin{itemize}
 \item Chances are that this CA model is also Turing complete, but it lacks a rigorous
       proof at present.
 \item Does there exist a spaceship translating itself across the grid
       horizontally or vertically?
 \item Does there exist a gun that can repeatedly shoot out gliders?
 \item Do there exist solitons that remain intact after an interaction with others?
 \item Can we carry out more experimentation of physics in this universe, such as
       percolation, quantum computing and lattice scattering?
 \item Discuss its mathematical structure in a strict form and connections to the
       theory of dynamical systems.
 \item Explore possible applications to signal encoding, random sequence generating,
       and even artificial intelligence.
 \item Generalize this model to higher dimensions, or stochastic automata.
\end{itemize}

\section{Conclusion\label{conclusion}}
 In this paper, we have constructed a new type of automata and discussed its links to
 some physical concepts. Much to our delight, the model also achieves a bewildering
 variety of patterns, especially dozens of easily controlled modules. Comparisons
 with Life are emphasized, and challenging problems are listed, but more simulations
 are not strived for. As has been revealed, the mathematical constructor is of
 simplicity, but no further discussions are provided.

 In closing, we would like to quote Zuse's perceptive predictions made forty years
 ago: \emph{Incorporation of the concepts of information and the automaton theory
 in physical observations will become even more critical, as even more use is made
 of whole numbers, discrete states and the like}~\cite{Zuse}. It is somewhat like
 Edwin Abbott Abbott's \textsl{Flatland}, which disseminated a crazy idea of many
 dimensions \emph{thereby contributing to the enlargement of the imagination}. A
 century later, string theorists have made great advances in their quest to the
 Holy Grail of physics~\cite{Witten, Mohaupt}. One of its underpinnings is just
 the assumption of 11 dimensions.

\addcontentsline{toc}{section}{References}
\phantomsection
\def\refname{\large{REFERENCES}}

\end{document}